\documentclass[conference]{IEEEtran}
\usepackage{graphicx}
\usepackage{tabularx}
\usepackage{times}
\usepackage{pifont}
\usepackage{amsmath}
\usepackage{amssymb}
\usepackage{verbatim}
\usepackage{moreverb}
\usepackage{alltt}
\usepackage{cite}
\usepackage{epsfig}
\usepackage{subfigure}
\usepackage{url}
\usepackage{multirow}
\usepackage{times,cite,subfigure,graphicx,epsfig,wrapfig,amsfonts,amssymb,algorithmic,threeparttable}

\IEEEoverridecommandlockouts

\begin{document}
\title{Secure Key Establishment for Device-to-Device Communications}

\author{\IEEEauthorblockN{{Wenlong Shen\IEEEauthorrefmark{1},
Weisheng Hong\IEEEauthorrefmark{1}, Xianghui Cao\IEEEauthorrefmark{1}, Bo Yin\IEEEauthorrefmark{1}, Devu Manikantan Shila\IEEEauthorrefmark{2} and
Yu Cheng\IEEEauthorrefmark{1}}
\IEEEauthorblockA{\IEEEauthorrefmark{1}Department of Electrical and
Computer Engineering, Illinois Institute of Technology, USA
\\Email: \{wshen7,whong4,byin\}@hawk.iit.edu; \{xcao10,cheng\}@iit.edu
}
\IEEEauthorblockA{\IEEEauthorrefmark{2}United Technologies Research Center. Email: manikad@utrc.com}
}}


\maketitle

\begin{abstract}

With the rapid growth of smartphone and tablet users, Device-to-Device (D2D) communications have become an attractive solution for enhancing the performance of traditional cellular networks. However, relevant security issues involved in D2D communications have not been addressed yet. In this paper, we investigate the security requirements and challenges for D2D communications, and present a secure and efficient key agreement protocol, which enables two mobile devices to establish a shared secret key for D2D communications without prior knowledge. Our approach is based on the Diffie-Hellman key agreement protocol and commitment schemes. Compared to previous work, our proposed protocol introduces less communication and computation overhead. We present the design details and security analysis of the proposed protocol. We also integrate our proposed protocol into the existing Wi-Fi Direct protocol, and implement it using Android smartphones.

\end{abstract}


\IEEEpeerreviewmaketitle

\section{Introduction}

The emergence and popularity of personal mobile devices, such as smartphones and tablets, generates large amount of data traffic by accessing the Internet and downloading applications, which imposes a huge burden for the cellular infrastructure and spectrum. Device-to-Device (D2D) communications have been introduced to offload the traffic burden from cellular infrastructure to personal devices \cite{d2d}. The D2D technology enables mobile device users directly establish wireless links between each other, without passing through the public cellular infrastructure or access points. 

Many literatures have studied the application scenarios and possible technical solutions for D2D communications. In \cite{d2dlte}, the authors propose D2D communications as an underlay to the cellular network, and present a mechanism for integrating D2D communications into LTE-Advanced network. Yu et al. \cite{d2dp, d2dp2} discuss the power control issue for D2D communications, and derive an optimum power allocation for D2D links under cellular network control. The work in \cite{5} proposes to use Wi-Fi based D2D links among cellular users to improve the overall network performance in uplink transmission. 

Wi-Fi Direct, initially called Wi-Fi P2P, is a Wi-Fi standard that enables devices to easily establish D2D connections using the Wi-Fi frequency band. \cite{wifidirect} gives a wide overview and experimental evaluation of the Wi-Fi Direct protocol. \cite{wifid2d} considers the practical implementation challenges of Wi-Fi Direct and shows that the Wi-Fi Direct features allow deploying the D2D paradigm on top of the LTE cellular infrastructure.

Though D2D communication has been a hot research topic in recent years, there is not much study focusing on the security aspect of D2D communications. \cite{phy1} and \cite{phy2} discuss the physical layer solutions for secure D2D communications, but their techniques are difficult to be implemented using devices on the market. 

In fact, due to the broadcast nature of wireless communication, wireless channels are considered vulnerable to a variety of attacks, and security is one of the major concerns for D2D communications. To secure the communication between two end users of a D2D link, establishing a shared secret key is the first and most significant step. However, lack of trusted third party and infrastructure under D2D connection environment makes this step a non-trivial task. The well-known Diffie-Hellman key agreement protocol enables two parties jointly establish a shared secret key without any prior knowledge. However, this protocol is vulnerable to the man-in-the-middle attack (MITMA)\cite{mao}: an active adversary makes independent connections with the victims, making them believe that they are talking directly to each other. To address this issue, researchers have come up with various Diffie-Hellman based cryptographic protocols, which can prevent the MITMA by conducting mutual authentication.

One simple protocol was suggested in \cite{tts}, in which devices A and B exchange the hashes of their public keys over a secure channel, thus performing the mutual authentication. However, this protocol requires a large number of bits to be mutually authenticated. The MANA protocol in \cite{mana} reduces the size of the authentication message to $k$ bits, but requires a stronger notation of authentication channel. \cite{ka} presents a protocol based on commitment schemes and requires 4-round communication over the wireless channel. In this paper, we propose a 3-round key agreement protocol based on commitment scheme. Our proposed protocol is similar to the protocol in \cite{ka}, but with less communication and computation overhead, meantime achieving the same level of security. Major contributions of this paper are summarized as follows:
\begin{enumerate}
  \item We analyze the secure threats and challenges for D2D communications;
  \item We design a secure and efficient Diffie-Hellman based key agreement protocol, and provide the security analysis;
  \item We integrate our proposed key agreement protocol into the existing Wi-Fi Direct protocol, and implement it on Android smartphones.
\end{enumerate}

The rest of this paper is organized as follows: Section II introduces the security concerns and challenges for D2D communications; Section III presents the details of the protocol design and security analysis; Section IV shows the implementation of our proposed protocol. We conclude this paper in Section V.

\section{Security Concerns of D2D Communication \label{sec2}}
Despite all the benefits of D2D communications, security is one of the major concerns that need to be well addressed before D2D technique gets widely accepted and implemented. It is well known that due to the broadcast nature of wireless channels, wireless communication such as Wi-Fi and Bluetooth is vulnerable to a variety of attacks that challenges the three basic principles of security--confidentiality, integrity and availability. Some common attack vectors include surreptitious eavesdropping, message modification and node impersonation. For example, by stealthy listening to the communication between two devices, an attacker can gain critical or privacy information, such as trade secrets or identity related information. Thus, the D2D communications between devices need to be properly secured. 

\begin{figure*}[ht]
  \centering
  \scalebox{0.7}
  {\includegraphics{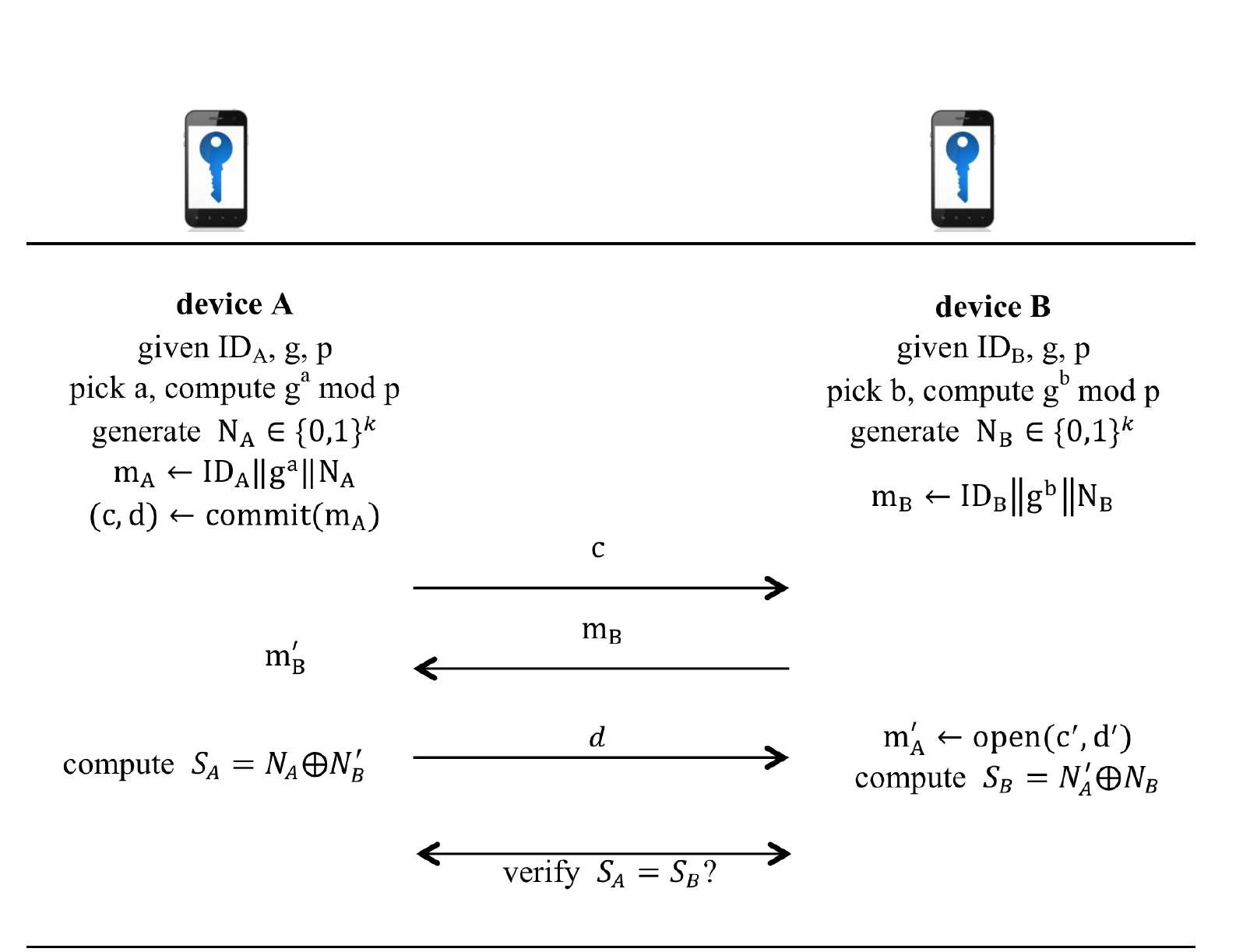}}
  \caption{Secure Key Exchange Protocol}
  \label{fig:protocol}
    \vspace{-3mm}
\end{figure*}

To secure the D2D communications, cryptography solutions are needed to encrypt the messages while they are transmitted via wireless channels. Numerous encryption algorithms have been well developed which can provide different security levels for the encrypted messages, but all of them require two devices agree on a shared secret (either a shared secret key or each other's public keys). Due to the large number of mobile devices, the diversity of device manufacturers and lack of standards, preloading secure keys into mobile devices is neither efficient nor practical. On the other hand, a trusted third party or infrastructure is not likely to be available in the D2D mobile environment. Thus, how to establish a shared secret between devices is one of the main challenges for secure D2D communications.

One straightforward way to establish a shared secret between two devices is that the two end users of the D2D link interactively set up a secret key via human negotiation (such as making a phone call if they are in distance). The problem for this is that the shared secret established by human interaction will be too weak in most cases. The attackers do not even need to be smart to crack this weak secret via brute force method, considering current computation power. To deal with this issue, cryptologists and researchers come up with two types of approaches which enable two individuals to establish a secure enough secret key: Diffie-Hellman key establishment protocol and secret key extraction from physical channel characteristics.

Physical layer based secret key generation methods have been proposed in recent years as alternative solutions for traditional Diffie-Hellman key agreement protocol. Unlike Diffie-Hellman key agreement protocol, whose security is guaranteed by the computational hardness of discrete logarithms, these physical layer based methods rely on the randomness and uniqueness of wireless fading channel properties: temporal variation, spatial variation and reciprocity. Generally, the two devices first send channel probing packets to measure the physical metrics of the wireless channel, then after using quantization and error correction technique, these two devices can yield the same secret key. The main problem for this type of methods is that the secret key generation rate is in most case very low. Users have to send lots of channel probing packets to achieve a secret key with enough bits and randomness. The communication overhead and relatively longer key generation time are not quite desirable for the case of D2D communications.

Diffie-Hellman cryptosystem is the oldest public key system still in use, which allows two individuals to agree on a shared secret key, even though they can only exchange messages over public channels. Diffie-Hellman key agreement protocol works as follows: Assume $p$ and $q$ are publicly known to two devices A and B (if not, A can put them into its message and send it to B), A and B both randomly generate a value $a$ and $b$. A computes $g^a$ mod $p$ and sends it to B, correspondingly, B computes $g^b$ mod $p$ then sends it to A. At the last stage, A computes $s=(g^a)^b$ mod $p$, B computes $s=(g^a)^b$ mod $p$. Both A and B will arrive at the same value, since $(g^a)^b$ and $(g^a)^b$ are equal mod $p$. $(g^a)^b$ mod p will be the established shared secret between A and B, thus can be subsequently used as encryption key for future communication. The implementation of Diffie-Hellman key agreement protocol requires some extent of computation capacity, since $p$, $a$, and $b$ can be quite large numbers. However, mainstream mobile devices on today's market have achieved gigahertz level processor frequency, so generating a secure enough shared secret, say, 156 bits, can be conducted within seconds.

As is well known, the above Diffie-Hellman key agreement protocol is vulnerable to the so called the man-in-the-middle attack. Since $g^a$ and $g^b$ are transmitted over the public channel, there is no way for device A to know for sure whether $g^b$ comes from device B, vice versa. Devices A will establish a shared secret with whoever transmits $g^b$, and it certainly might not be device B. The essential reason that the MITMA is possible is that there is no mutual authentication between these two devices. To provide the desired authentication, one intuitive solution is both devices put the obtained secret key to a one-way hash function, e.g. MD5, to generate a hash value h(K), then compare the hash value via a trusted channel (for example, output the computed hash code on device screens and perform visual or verbal comparison). If the mutual authentication process agrees, then both devices can confirm that they have established a shared secret key with each other.

The main issue about the above mutual authentication procedure is that the number of bits needed to be checked by the user is too large. The output of a hash function is usually over 128 bits (32 hexadecimal digits), and visually or verbally checking them is a non-trivial task. Using truncation of the hash code can drastically reduce the number of digits needed to be checked, but doing this will introduce serious security weakness. In \cite{mana}, the authors describe one possible way to attack the truncated hash code: an attacker with significant computing resources can crack a 32 bits truncated hash code in less than 1 second.
Through the analysis above, to secure the D2D communications, we need a key agreement protocol that enables two mobile devices to securely establish a shared secret key, at the meantime requires minimum amount of information to be mutually authenticated to prevent the MITMA.

\section{Proposed Key Exchange Protocol \label{sec3}}


\subsection{Problem Statement}

We consider the following scenario. Two mobile device users want to establish a shared secret key for their D2D communications. Both of them are equipped with a smartphone or tablet which is capable of communicating over a wireless channel. Both devices have the computation capacity to perform Diffie-Hellman key agreement protocol, and are capable of displaying sequence of digits. The two users do not have any pre-shared cryptographic information, and there is no trusted third party or infrastructure available. They can visually or verbally recognize each other for the purpose of mutually authenticate a short message.

\subsection{Assumptions}

We assume devices A and B agree on a finite cyclic group $\mathbb{G}$, its generating element $g$, and a large prime number $p$. We assume $\mathbb{G}$ to be a subgroup of $\mathbb{Z}_p^*$ of prime order q, where, $\mathbb{Z}_p^*$ is the multiplicative group consists of nonzero integers modulo $p$.

We consider the Dolev-Yao adversary model \cite{mao}: The attacker has fully control over the wireless channel. It can overhear, intercept, and modify any message. The attacker can also initiate a conversation with any other user. We further assume that legitimate users will follow the protocol and are not compromised.

\subsection{Commitment Schemes}
A commitment scheme allows one user to commit to a chosen value or statement while keep it hidden to others, with the ability to reveal the commitment value latter. A commitment scheme has the following two main properties: 1) a user cannot modify the value or statement after they have committed to it; that is, the commitment scheme is binding and 2) the receiver can only know the committed value after the sender ``opens" it; that is, the commitment scheme is hiding. A commitment scheme is defined by two algorithms \textbf{Commit} and \textbf{Open}:

\noindent $\textbf{Commit}.\;(c, d)\leftarrow m$ transforms a value $m$ into a commitment/open pair $(c, d)$. The commit value c reveals no information of m, but with decommit value $d$ together $(c, d)$ will reveal $m$.

\noindent $\textbf{Open}.\;m\leftarrow (c, d)$ output original value $m$ if $(c, d)$ is the commitment/open pare generated by $\textbf{Commit}(m)$.

\subsection{Protocol Design}

Here we present our design of the key agreement protocol, which is based on the traditional Diffie-Hellman key agreement protocol and a commitment scheme. In out protocol, two mobile users A and B respectively generate $k$-bit random strings $\mathrm{N}_A$ and $\mathrm{N}_B$, and $\mathrm{N}_A\oplus \mathrm{N}_B$ as the short authentication string for mutual authentication.

Fig. \ref{fig:protocol} shows the message flow of our proposed protocol. At the initial stage, user A and B select their Diffie-Hellman parameter $a$ and $b$, then compute $g^a$ and $g^b$. A and B randomly generate their $k$-bit strings $\mathrm{N}_A$ and $\mathrm{N}_B$. $m_A= \mathrm{ID}_A\lVert g^a\rVert \mathrm{N}_A$ and $m_B= \mathrm{ID}_B\lVert g^b\rVert \mathrm{N}_B$ are formed by concatenation, in which $\mathrm{ID}_A$ and $\mathrm{ID}_B$ are human readable identifiers for user A and B, such as names or e-mail addresses. A also needs to calculates the commitment/opening $(c, d)$ for $m_A= \mathrm{ID}_A\lVert g^a\rVert \mathrm{N}_A$. 

After the initial stage, user A and user B perform the following message exchange over their D2D communications channel. User A sends the $c$, the commitment value of $m_A$ to user B; after receiving $c$, user B sends $m_B$ to user A. In return, user A sends the decommit value $d$ to user B. User B opens the commitment and gets $m_A= \mathrm{ID}_A\lVert g^a\rVert \mathrm{N}_A$. 

In the final stage, user A and B generate the k bits authentication string by $S_A=\mathrm{N}_A\oplus \mathrm{N}_B'$ and $S_B=\mathrm{N}_A'\oplus \mathrm{N}_B$, in which $\mathrm{N}_B'$ and $\mathrm{N}_A'$ are derived from messages received by A and B. Then user A and B verify if $S_A=S_B$ via trusted channel (visual or verbal comparison). If the authentication strings match, A and B accept each other's Diffie-Hellman parameters and calculate the shared secret key $K=g^{ab} \;\mathrm{mod}\; p$. The reason for comparing authentication string before generating Diffie-Hellman secret key is that if the strings do not match, both users can save the computation for secret key generation.

\begin{figure*}[htbp]
  \centering
  \scalebox{0.571}
  {\includegraphics{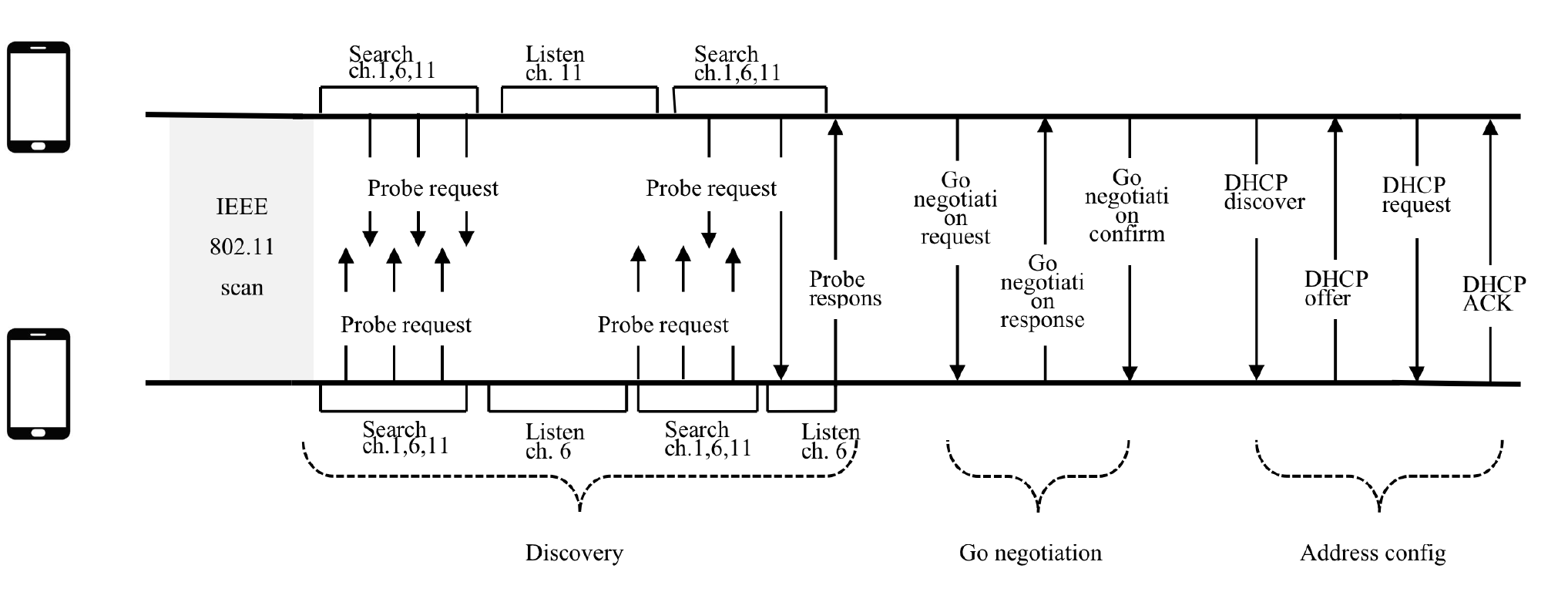}}
  \caption{Wi-Fi Direct Protocol}
  \label{fig:fig2}
    \vspace{-3mm}
\end{figure*}

\begin{figure*}[htbp]
  \centering
  \scalebox{0.571}
  {\includegraphics{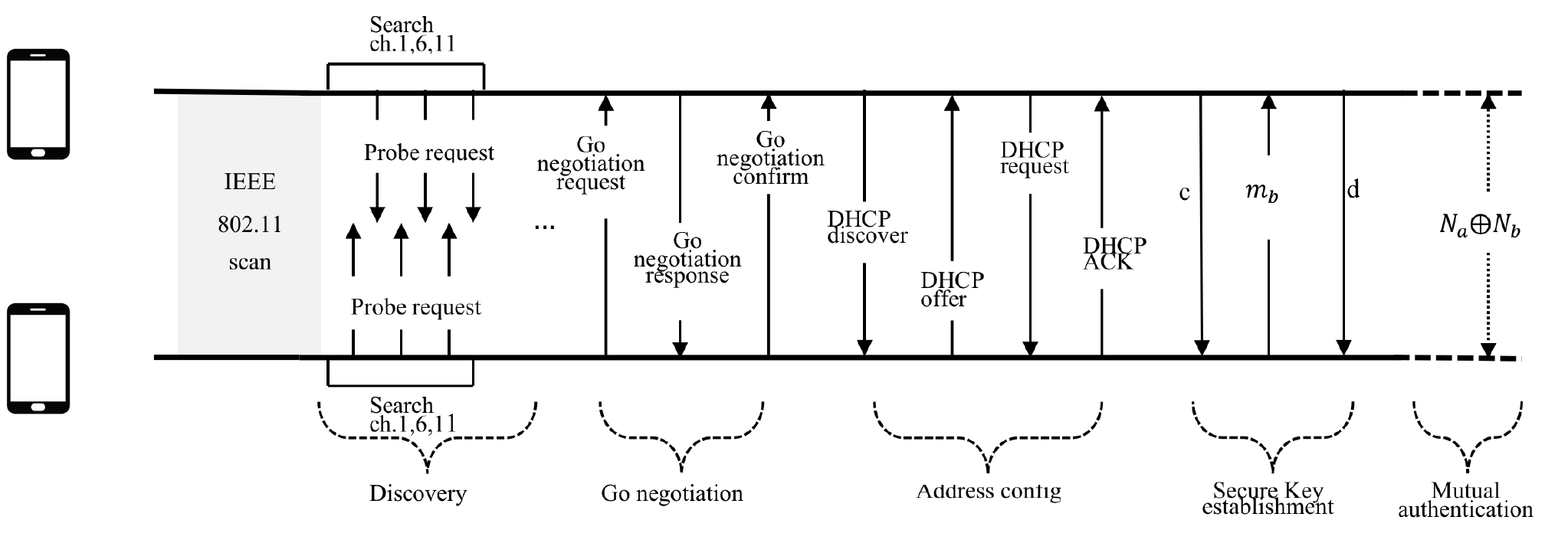}}
  \caption{Secure Wi-Fi Direct Protocol}
  \label{fig:fig3}
  \vspace{-3mm}
\end{figure*}

\subsection{Security Analysis}
In our security analysis, we assume the commitment scheme we use to be an ideal commitment scheme. That is, no attacker can forge an $m'$ which yields a commitment value $c'$ such that $c'=c$ ($c$ is the commitment value of original message $m$); and no attacker can open a commitment with $d'\neq d$. We further assume that both $\mathrm{N_A}$ and $\mathrm{N_B}$ generated by devices A and B are perfectly random. 

Notice that in our proposed protocol, any party has to commit on an $m_A'$ before actually seeing $m_B$; and any party has to submit an $m_B'$ before actually seeing $m_A$. These statements directly follow from the binding and hiding properties of the commitment scheme. Thus, no matter what attacking strategy the attacker applies, it has to first commit to or submit its own $m$ message. Suppose the attacker E initiate a protocol with user B pretending itself to be A, it will first commit to an $m_E= \mathrm{ID}_A\lVert g^e\rVert \mathrm{N}_E$ and send the commit value $c_E$ to B. After getting the reply message $m_B$, the attacker can modify $m_B$ into $m_B'= \mathrm{ID}_B\lVert g^e\rVert \mathrm{N}_B$ and forward it to A. But when it comes to the final stage of the protocol, A and B will compare $S_B=\mathrm{N}_B\oplus\mathrm{N}_E$ with $S_A=\mathrm{N}_A\oplus\mathrm{N}_B$. The only chance that A and B agree on the authentication string is $\mathrm{N}_E=\mathrm{N}_A$. Due to the binding property of the commitment scheme, the attacker E cannot modify $\mathrm{N}_E$ after it sends out its commitment. The probability that E launch a successful attack is at most $2^{-k}$ ($k$ is the number of bits of the authentication string). If the attacker launch an attack by replying $M_E= \mathrm{ID}_B\lVert g^e\rVert \mathrm{N}_E$ to a protocol initiator A, similar analysis follows. 

The bit length $k$ of authentication string can be tuned to balance the trade-off between security level and usability. With a larger $k$, users gain higher security level but need to compare a longer authentication string. Usually we consider 20 bits (5 hexadecimal digits) to be securely enough, this will give us a security level roughly equal to an ATM machine.

\section{Protocol Implementation \label{sec4}}

In this section, we integrate our proposed key agreement protocol into the existing Wi-Fi Direct protocol. We call this enhanced version ``Secure Wi-Fi Direct" protocol, which provides secure key establishment functionality for the D2D connection  between two mobile devices. We also implement the secure Wi-Fi Direct protocol on two Android smartphones. The implementation result shows that these two phones obtain a shared secret key at the meantime they establish a D2D connection. 

\begin{figure}[htbp]
  \centering
  \scalebox{0.4}
  {\includegraphics{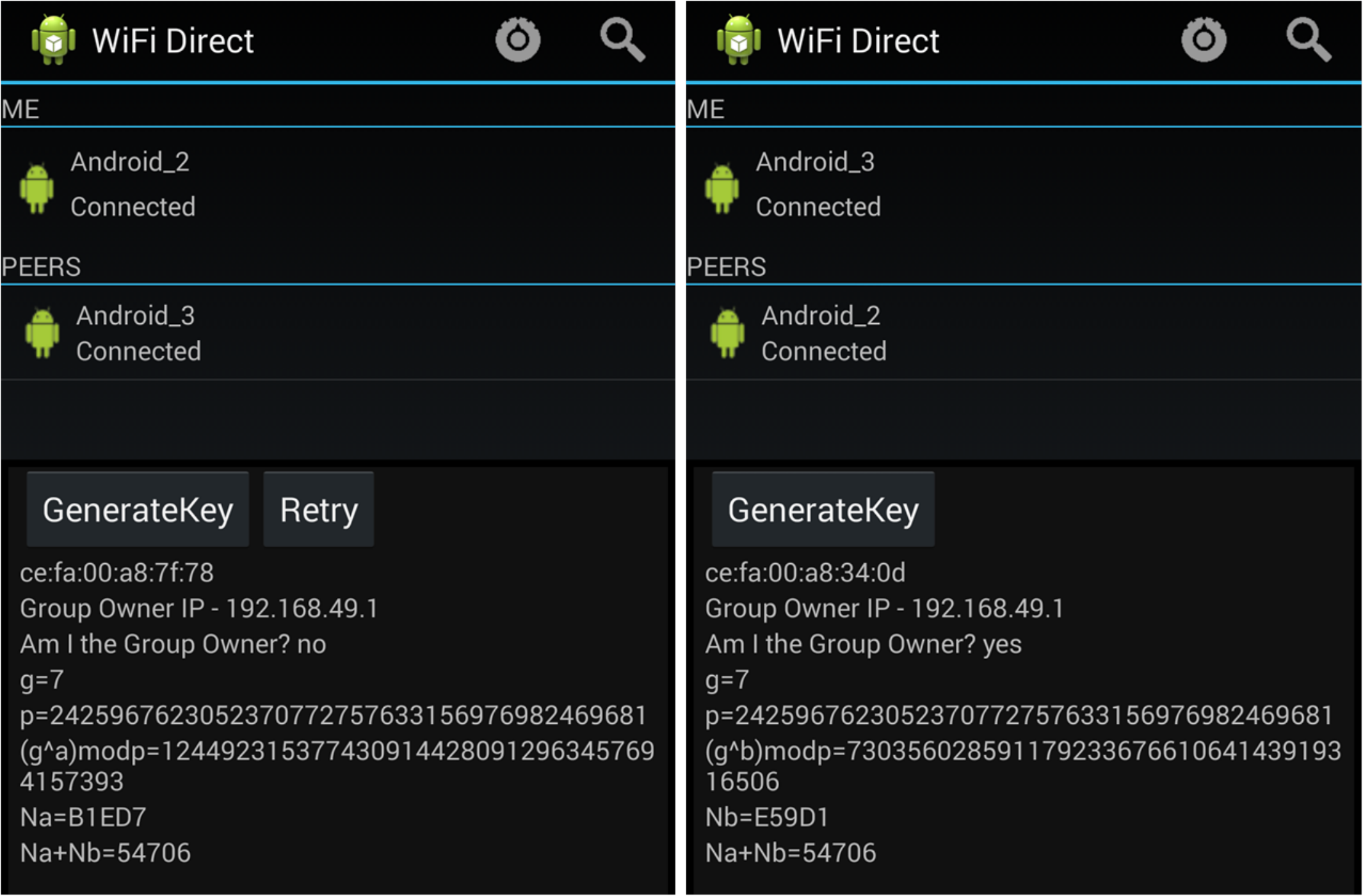}}
  \caption{Experiment Results}
  \label{fig:fig1}
\end{figure}

Wi-Fi Direct protocol enables two devices to establish a D2D connection using Wi-Fi frequency without the help of access points. Fig. \ref{fig:fig2} shows the procedure for a D2D connection establishment using Wi-Fi Direct. First, two devices perform the channel probing and discover each other. Then the two devices will go through a 3 way handshake to determine the group owner (works as an access point) for this D2D connection. After the devices have agreed on their respective roles, a DHCP exchange will be conducted to set up the IP addresses for both devices. Thus, the D2D connection between these two devices has been established. We add our proposed key agreement protocol on top of the existing Wi-Fi Direct protocol, as is shown in Fig. \ref{fig:fig3}. After the address configuring phase, the two devices will go through our proposed key agreement protocol as well as the mutual authentication process to agree on a shared secrete key. As long as the two devices have agreed on the authentication message, they can subsequently use their shared secret key for future communication.

We implement the secure Wi-Fi Direct protocol on two Android smartphones. The model and operation system of the smartphones we use is Nexus 5 and the newest AndroidOS-4.4 KitKat. We build our application based on the Wi-Fi Direct Demo application in \cite{demo}. Our secure protocol is implemented by programming the Android TCP socket.  The result is shown in Fig. \ref{fig:fig1}. We list some of the Diffie-Hellman parameters, as well as authentication $\mathrm{N}$ strings in hexadecimal. We use a 40 digits $p$ value, which gives us a roughly 130 bits secret key. The authentication string length is set to be 20 bits (5 hexadecimal digits), which is easy to be compared by the two users and can achieve a strong security level. The overall running time for our security protocol, excluding the user-conducted mutual authentication part, including the computation time as well as the communication delay, is trivial on Nexus 5 smartphone with its 2.26 GHz processor.

\section{Conclusion \label{sec5}}

In this paper, we analyzed the secure requirements and challenges for secret key establishment between two mobile devices. The proposed key agreement protocol enables two mobile users to securely set up a secret key with a small computation cost and low mutual authentication overhead. The security analysis of the proposed protocol shows that the probability for an attacker to launch a successful attack is at most $2^{-k}$, where $k$ is the number of bits used for authentication strings. We also integrated our key agreement protocol into the existing Wi-Fi Direct protocol, and implemented it using real smartphones. The implementation result shows that our proposed protocol is efficient, and achieves high level of usability.
 
\section*{Acknowledgment}
This work was supported in part by the NSF under grants CNS-1320736 and CNS-1117687.


\begin{thebibliography}{99}

\bibitem{d2dlte}
K. Doppler, M. Rinne, C. Wijting, C.B. Ribeiro, and K. Hugl, ``Device-to-device communication as an underlay to LTE-advanced networks,''  \emph{IEEE Communications Magazine}, vol. 47, no. 12, pp. 42-49, 2009.

\bibitem{d2dp}
C. Yu, O. Tirkkonen, K. Doppler, and C. Ribeiro, ``Power optimization of device-to-device communication underlaying cellular communication,'' in \emph{Proc. IEEE ICC}, pp. 1-5, 2009.

\bibitem{d2dp2}
C. Yu, K. Doppler, C.B. Ribeiro, and O. Tirkkonen, ``Resource sharing optimization for device-to-device communication underlaying cellular networks,'' \emph{IEEE Trans. Wireless Commun.}, vol. 10, no. 8, pp. 2752-2763, 2011.


\bibitem{5}
A. Asadi and V. Mancuso, ``Energy efficient opportunistic uplink packet forwarding in hybrid wireless networks,'' in \emph{Proceedings of the fourth international conference on future energy systems}, ACM pp. 261-262, 2013

\bibitem{wifidirect}
A.G. Saavedra and P. Serrano, ``Device-to-device communications with WiFi Direct: overview and experimentation,'' \emph{IEEE Wireless Communications}, vol. 20, no. 3, 2013.

\bibitem{wifid2d}
A. Asadi and V. Mancuso, ``WiFi Direct and LTE D2D in action,'' \emph{Wireless Days (WD)}, 2013.

\bibitem{tts}
D. Balfanz, D.K. Smetters, P. Stewart, and H.C. Wong, ``Talking to strangers: authentication in Ad-Hoc wireless networks,'' in \emph{Proc. Network and Distributed System Security Symposium Conference}, 2002.

\bibitem{mana}
C. Gehrmann, C.J. Mitchell, and K. Nyberg, ``Manual authentication for wireless devices,'' \emph{RSA Cryptobytes}, vol. 7, No. 1, pp. 29-37, 2004.

\bibitem{ka}
M. Cagalj, S. Capkun, and J.P. Hubaux, ``Key agreement in peer-to-peer wireless networks,'' in \emph{Proc. IEEE (Special Issue on Cryptography and Security)}, 2006.

\bibitem{phy1}
J. Wang, Ch. Li, and J. Wu, ``Physical layer security of D2D communications underlaying cellular networks,'' \emph{Applied Mechanics and Materials}, vol. 441, pp. 951-954, 2014.

\bibitem{phy2}
D. Zhu, A.L. Swindlehurst, S.A. Fakoorian, W. Xu, and Ch. Zhao, ``Device-to-device communications: the physical layer security advantage.'' in \emph{IEEE ICASSP}, 2014.

\bibitem{mao}
W. Mao, \emph{Modern Cryptography: Theory and Practice}, Prentice Hall PTR, New Jersey, USA, 2004

\bibitem{demo}
Wi-Fi Direct Demo, available on line: http://www.androidside.com

\bibitem{d2d}
G. Fodor, E. Dahlman, G. Mildh, S. Parkvall, N. Reider, G. Miklos, and Z. Turanyi, ``Design aspects of network assisted device-to-device communications,'' \emph{IEEE Communications Magazine}, vol. 50, no. 3, pp. 170-177, 2012.





\end{thebibliography}

\bibliographystyle{ieeetr}

\end{document}